\documentclass[10pt,letterpaper,twocolumn]{article} 

\usepackage{ol2}
\usepackage[draft]{hyperref}
\usepackage{amsmath}

\begin{document}

\twocolumn[ 

\title{Supersymmetric transparent optical intersections}


\author{Stefano Longhi}

\address{Dipartimento di Fisica, Politecnico di Milano and Istituto di Fotonica e Nanotecnologie del Consiglio Nazionale delle Ricerche, Piazza L. da Vinci 32, I-20133 Milano, Italy (stefano.longhi@polimi.it)}

\begin{abstract}
Supersymmetric (SUSY) optical structures provide a versatile platform to manipulate the scattering and localization properties of light, with potential applications to 
mode conversion, spatial multiplexing and invisible devices. Here we show that SUSY can be exploited to realize broadband transparent intersections
between guiding structures in optical networks for both continuous and discretized light. These include transparent crossing of high-contrast-index waveguides and directional couplers, as well as crossing of guiding channels in coupled resonator lattices. 
\end{abstract}

\ocis{230.7370, 290.0290, 260.2710, 000.1600}


 ] 

\noindent

The synthesis of optical structures with desired scattering properties is of major importance for a wide variety of applications. In the past decade, novel powerful tools of inverse scattering, such as those based on conformal mapping and transformation optics (TO)  \cite{TO1,TO2}, have been introduced, leading to the design and realization of novel devices  such as invisible cloaks, illusion objects, field concentrators, and perfect 'black hole' absorbers \cite{TO3,TO4,TO5,TO6,TO7}. Recently, a synthesis method based on the optical analogue of
supersymmetry (SUSY) has been introduced \cite{S1,S2,S3}. SUSY optical structures display several interesting properties
with potential applications to global phase matching, efficient mode conversion and fully-integrated spatial multiplexing \cite{S1,S4}. 
SUSY optical structures also enable to realize transparent defects and interfaces \cite{S5,S6,S7}. As compared to TO methods, SUSY  shows less stringent requirements of  material parameters \cite{S1,S2} and can be applied to discretized light in coupled waveguide or resonator structures as well \cite{S4,S5,S8}.\\ In this Letter the potentialities of optical SUSY for the design of transparent intersections in integrated optical networks are disclosed. The ability to efficiently intersect high index contrast optical waveguides with little or no signal deterioration is crucial in constructing high-density integrated optical circuits. Owing to waveguide crossing, an optical signal typically experiences scattering, both into radiation modes and to guided modes, generating a detrimental back-reflected wave and crosstalk. Several methods have been proposed and demonstrated to reduce back-reflection and crosstalk at the intersections between two dielectric waveguides, including multimode interference
 structures \cite{WC1,WC2}, resonant coupling \cite{WC3},  elliptical or parabolic
mode expanders \cite{WC4,WC5,WC6}, graded-index (GRIN) waveguides \cite{WC7}, and guiding top layers \cite{WC8}, to mention a few.
SUSY provides a natural platform to synthesize transparent  crossing of optical components. Here we show that broadband transparent intersections are possible  for high index contrast waveguides, as well as for more complex optical components such as directional couplers. Back-reflection-free crossing is also shown to occur for discretized light at the intersection of guiding channels in lattices of coupled resonators.
 \begin{figure}[htb]
\centerline{\includegraphics[width=8cm]{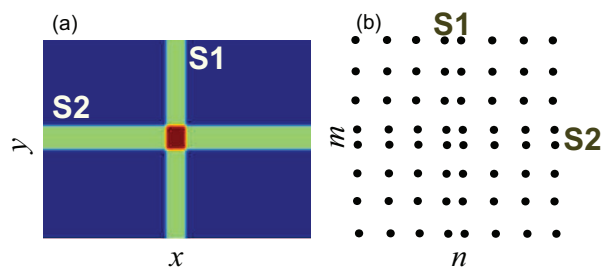}} \caption{ \small
(Color online) Schematic of orthogonal intersection between two guiding structures S1 and S2 in (a) a continuous 2D dielectric medium, and (b) in a square lattice of coupled resonators with defects.}
\end{figure}
We consider light propagation in a two-dimensional (2D) dielectric medium with a refractive index distribution $n(x,y)= \sqrt{\epsilon_r(x,y)}$, that describes the intersection of two guiding structures. We focus our analysis to TE-polarized waves $(E_x=E_y=H_z=0)$, which is more suited for the application of SUSY in a purely dielectric medium \cite{S2}. For a TE wave, the $E_z$ component of the electric field satisfies the Helmholtz equation 
\begin{equation}
\frac{\partial^2 E_z}{\partial x^2}+\frac{\partial^2 E_z}{\partial y^2}+\beta^2 n^2(x,y)E_x=0
\end{equation}   
where $x$ and $y$ are the spatial coordinates, normalized to a characteristic spatial length $a$ (defining e.g. the typical width of waveguides), $\beta=(\omega a/c_0)= 2 \pi a / \lambda$, and $\omega=2 \pi c_0 / \lambda$ is the frequency of the electromagnetic wave with (vacuum) wavelength $\lambda$. For an arbitrary distribution of the refractive index, SUSY can not be applied in a simple way, though SUSY extensions to the Helmholtz equation,  based on Moutard transformation, have been suggested \cite{Mou}. However, for a refractive index distribution of the form $n^2(x,y)=n_0^2+\Delta \epsilon_{rx}(x)+\Delta \epsilon_{ry}(y)$, separation of variables is possible, and standard SUSY of the 1D Schr\"{o}dinger equation can be exploited to engineer the scattering properties of Eq.(1). In the previous expression of $n^2(x,y)$, $n_0$ is a reference (substrate) refractive index, whereas $\Delta \epsilon_{rx}(x)$ and $\Delta \epsilon_{ry}(y)$ describe the dielectric profiles of the two guiding structures S1 and S2, respectively, that intersect each other at 90$^{\rm o}$, with $\Delta \epsilon_{rx}(x), \Delta \epsilon_{ry}(y) \rightarrow 0$ as $x,y \rightarrow \pm \infty$; see Fig.1(a). After setting $E_z(x,y)=E_{zx}(x)E_{zy}(y)$, Eq.(1) splits into the two stationary 1D Schr\"{o}dinger-type equations 
 \begin{figure}[htb]
\centerline{\includegraphics[width=8.6cm]{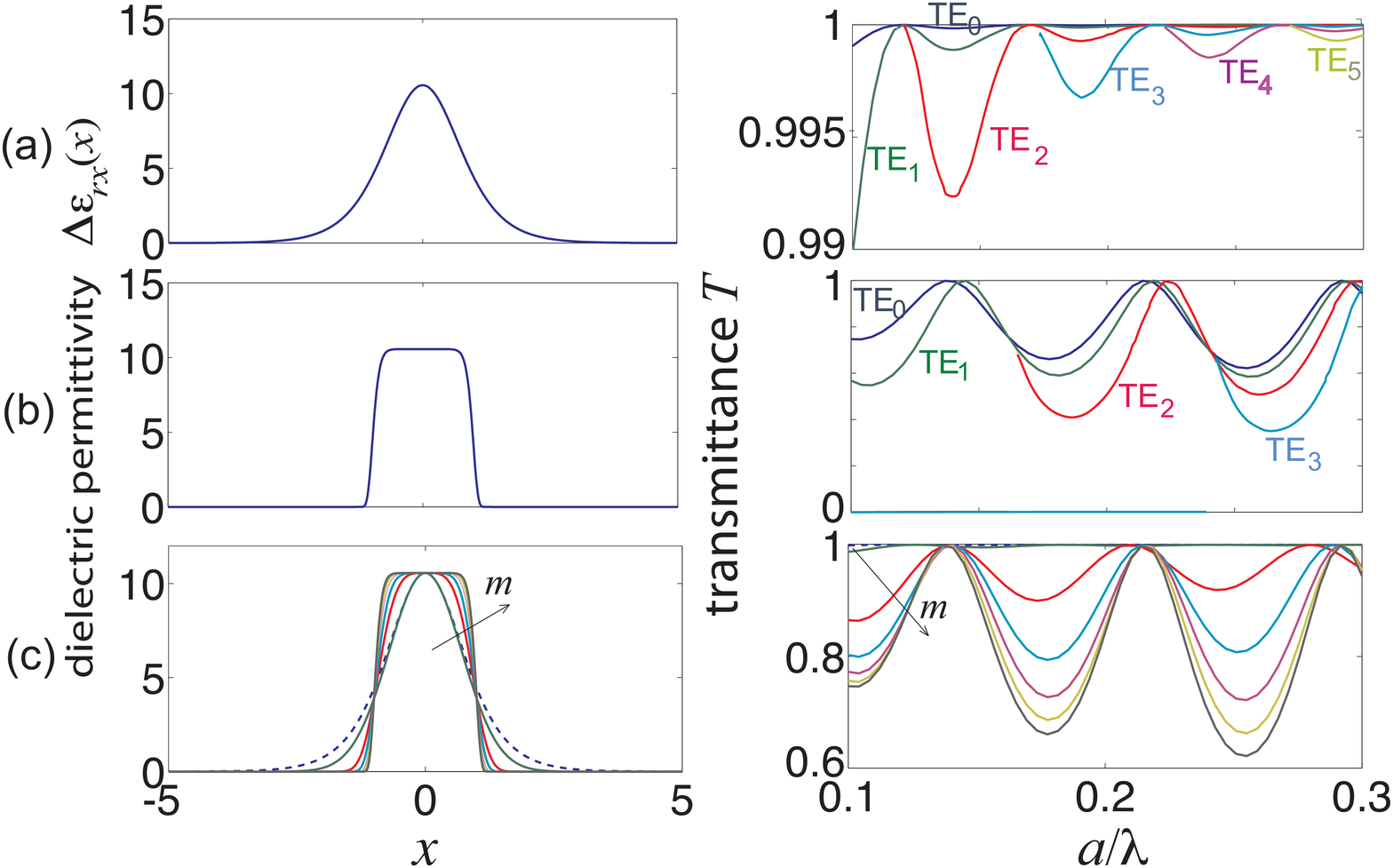}} \caption{ \small
(Color online) (a) Permittivity profile $\Delta \epsilon_{rx}$ of SUSY waveguide (left panel) and numerically-computed transmission spectra of the various TE-polarized modes (right panel). $T=1$ is exactly achieved at the normalized frequencies $a / \lambda=(1/ 2 \pi) \sqrt{l(l+1) / \Delta}$. (b) Same as (a), but for a waveguide with a super-Gaussian index profile of order $m=6$. (c) Permittivity profiles (left panel) and corresponding transmission spectra (right panel) of the fundamental TE$_0$ mode for a super-Gaussian profile at increasing order $m=1,2,3,4,5$ and $6$. The dotted curves refer to the SUSY waveguide in (a).}
\end{figure}
\begin{eqnarray}
\frac{d E_{zx}}{d x^2}+\beta^2 \Delta \epsilon_{rx}(x)E_{zx}=-\gamma_x E_{zx} \\
\frac{d E_{zy}}{d y^2}+\beta^2 \Delta \epsilon_{ry}(y)E_{zy}=-\gamma_y E_{zy} \; , \; 
\end{eqnarray}
where $\gamma_{x,y}$ are the separation constants, with $\beta^2n_0^2=\gamma_x+\gamma_y$. Owing to the factorization of $E_z$, cross-talk (i.e. light scattering into guide S1) is fully suppressed, {\it regardless} of the shapes of $\Delta \epsilon_{rx,ry}$, and the only scattering process that makes the crossing not transparent is back-reflection. If $\beta^2 \Delta \epsilon_{rx}(x)$ is a reflectionless potential, the guiding structure S1 is transparent and any {\it arbitrary} field distribution propagating in the guide S2 is not reflected at the intersection of guide S1. More precisely, let $u_y(y)$ be a guided mode of S2 with eigenvalue $\gamma_y=\gamma_{y0}$. Then, far from the crossing region, the solution to Eq.(1) can be written as $E_{z}(x,y) \sim u_y(y) [\exp(i \sqrt{\gamma_{x0}} x)+r \exp(-i \sqrt{ \gamma_{x0}}x)]$ as $x \rightarrow -\infty$, and $E_{z}(x,y) \sim u_y(y) t  \exp(i \sqrt{\gamma_{x0}}x)$ as $x \rightarrow \infty$, where $r$ and $t$ are the reflection and transmission coefficients of the potential $\beta^2 \Delta \epsilon_{rx}(x)$ for the Schr\"{o}dinger equation (2) with $\gamma_x \equiv \gamma_{x0}=\beta^2 n_0^2-\gamma_{y0}$. Note that, since the potential $\beta^2 \Delta \epsilon_{rx}(x)$ depends on the wavelength via $\beta$, strictly speaking a reflectionless potential -and thus exact transparent crossing- can be obtained at a prescribed wavelength. Nevertheless, numerical results show that the transmittance $T=|t|^2$ remains close to one in a broad wavelength range. As a first example, let us consider the crossing of two equal waveguides S1 and S2 with GRIN profiles belonging to the simplest family of reflectionless potentials obtained by first-order SUSY of a homogeneous medium, namely $\Delta \epsilon_{rx}(x)=\Delta {\rm sech}^2(x)$ and $\Delta \epsilon_{ry}=\Delta \epsilon_{rx}$. Note that $n_p=\sqrt{\Delta+n_0^2}$ determines the peak index change of the GRIN guide. The potential is strictly reflectionless when $\beta^2 \Delta=l(l+1)$, with $l=1,2,3,...$ \cite{uffa}. Figure 2(a) shows the transmittance $T$ versus the normalized frequency $\beta / ( 2 \pi)=a/ \lambda$  of the various TE-polarized guided modes for parameter values taken from Ref.\cite{WC3}, i.e. $n_0=1$ (air) and $n_p=3.4$ (GaAs), neglecting for the sake of simplicity the dependence of $n_p$ on wavelength. The transmittance $T$ has been numerically computed by a standard transfer matrix method from the 1D Schr\"{o}dinger equation (2), after computation of the propagation constants $\gamma_{y0}$ of the various guided modes from Eq.(3). Figure 2 clearly shows that high transmittance ($>99\%$) over a broad spectral range is observed for all guided modes of the structure. This is a very distinct and improved result as compared to e.g. the resonant tunneling method \cite{WC3}, where high transmittance and low crosstalk is obtained in a much  narrower spectral region (see Fig.5 of Ref.\cite{WC3}). 
GRIN profiles such as those required to realize SUSY transparent crossing  can be implemented in structured slab waveguides with subwavelength
holes etched in a waveguide \cite{ref1}. Using this technique, a
refractive index ranging from the bulk value of the dielectric host
medium to near unity is achievable \cite{ref1,ref2}. Alternatively, GRIN distributions can be implemented by controlling the thickness of the
guiding layer of a slab waveguide \cite{ref3}. 
Deviations of the GRIN profile from the reflectionless one causes a degradation of the transmittance. Figure 2(b) shows, as an example, the behavior of the transmittance $T$ computed for a super-Gaussian index profile $\Delta \epsilon_{rx}=\Delta \exp(-x^{2m})$ with $m=6$, i.e. corresponding to a nearly step-index guide. As the super-Gaussian order $m$ is decreased and the potential shape of reflectionless type is approximated, a clear improvement of the transmittance is observed, see Fig.2(c). The figure indicates that deviations from the exact SUSY profile due to fabrication imperfections can be partially tolerated; for example, for a Gaussian ($m=1$) rather than sech$^2$ profile a broadband transmittance larger than $98.5 \%$ is observed. The sech$^2$-like index profile is strictly transparent solely for TE-polarized waves, because a dielectric profile that is transparent to TE waves it is {\it not} for TM waves \cite{S2}. Nevertheless, numerical results based on 2D FDTD simulations of Maxwell's equations show that the sech$^2$-like index profile obtained by SUSY for TE-polarized modes yields negligible back reflection and crosstalk for TM waves as well; see as an example Fig.3.\\
An interesting property of SUSY is that almost transparent crossing over a broad frequency range can be realized for more complex structures than simple waveguides. For example, transparent crossing of two optical directional couplers, or of an optical directional coupler and a waveguide, can be designed. A transparent optical directional coupler with a desired coupling length can be synthesized by application of a double SUSY, starting from a homogeneous medium. Its profile is given by
\begin{equation}
\Delta \epsilon_{rx}(x)=\Delta \frac{\sigma^2+{\rm sech}^2(x) {\rm sinh}^2(\sigma x)}{\left[ {\rm tanh}(x) {\rm sinh}(\sigma x)-\sigma {\rm cosh}(\sigma x) \right]^2}
\end{equation}  
which is reflectionless for $\Delta \beta^2=2 (\sigma^2-1)$. In Eq.(4), the parameter $\sigma>1$ determines the coupling length between the guides of the coupler, which is given by $L=\pi/[(\sigma^2-1)]$. As an example, Fig.4(a) shows the transmittance of the coupler supermodes  at the intersection for $\Delta=3.941$ and $\sigma=1.2$. For comparison, the transmittance of directional couplers with a double-well super-Gaussian profile is depicted in Figs.4(b) and (c), showing a strong oscillating behavior with large back-reflectance for nearly step-index profile.

 \begin{figure}[htb]
\centerline{\includegraphics[width=8.6cm]{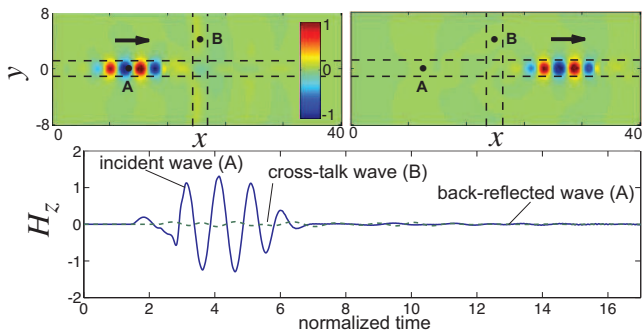}} \caption{ \small
(Color online)  2D FDTD  propagation of a TM-polarized wave packet at the crossing of two waveguides with a sech$^2$-like index profile [left panel of Fig.2(a)]. The carrier wavelength of the incident wave packet is $\lambda/a=10$. The upper plots show two snapshots of $H_z(x,y)$ (in arbitrary units) before (left panel) and after (right panel) the crossing. Arrows indicate the direction of propagation, whereas the straight dashed lines schematically depict the guiding regions. The lower plot shows the behavior of $H_z$ versus time, normalized to the optical period of oscillation, at the two points A (solid curve) and B (dashed curve). The signal in B corresponds to the crosstalk wave, whereas the delayed signal in A corresponds to the back-reflected wave.}
\end{figure}
   
 \begin{figure}[htb]
\centerline{\includegraphics[width=8.6cm]{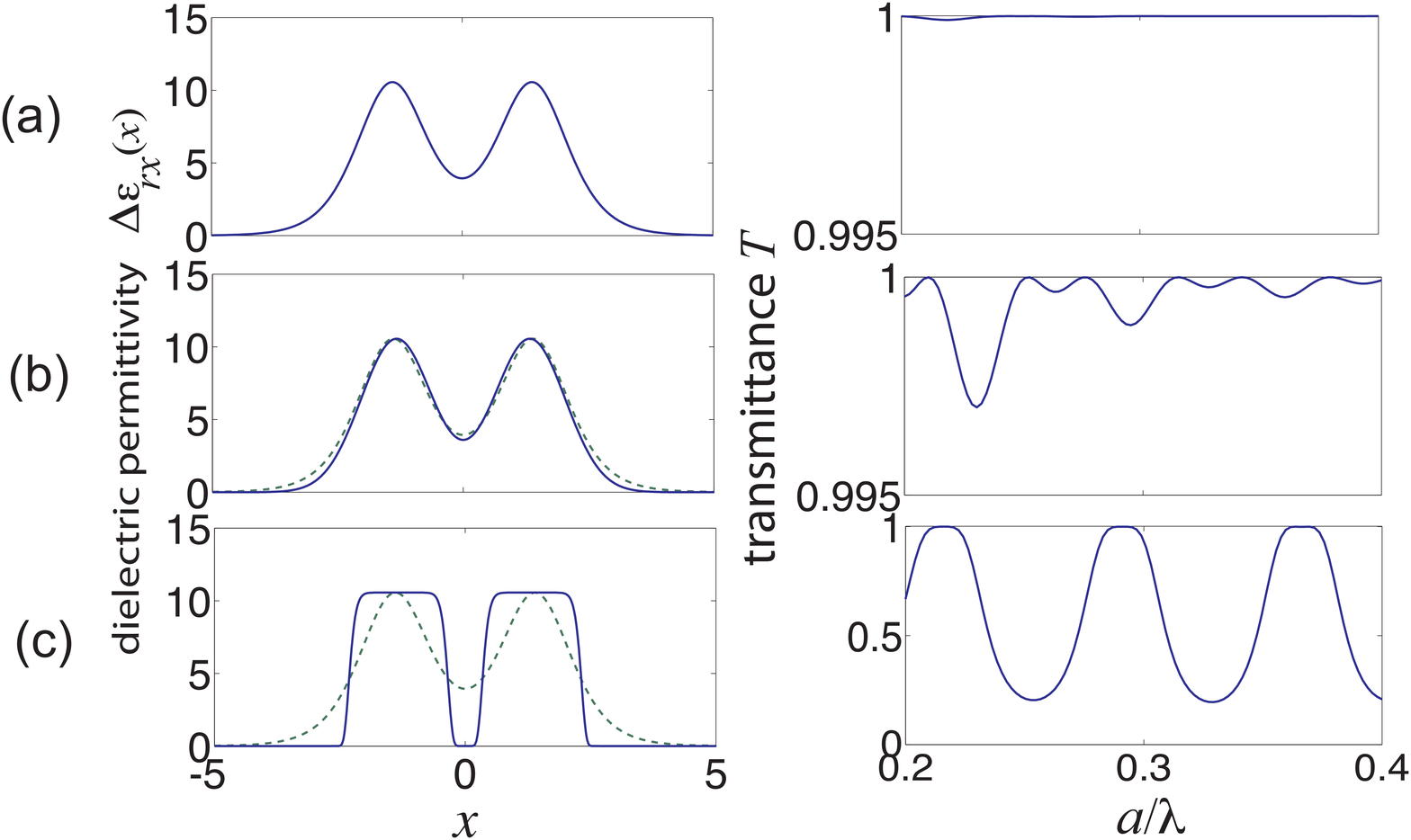}} \caption{ \small
(Color online)  (a) Transmission spectrum (right panel) of  the supermodes for the intersection of two SUSY-synthesized directional couplers [Eq.(4)]. Left panel: permittivity profile of the SUSY coupler $(n_0=1$, $\sigma=1.2$, $\Delta=3.941$). (b), (c):  Same as (a), but for a directional coupler made of two super-Gaussian guides of mode order $m=1$ [in (b)] and $m=6$ [in (c)]. The dotted curves in the left panels show the SUSY reference permittivity profile of (a).}
\end{figure}

 \begin{figure}[htb]
\centerline{\includegraphics[width=8.6cm]{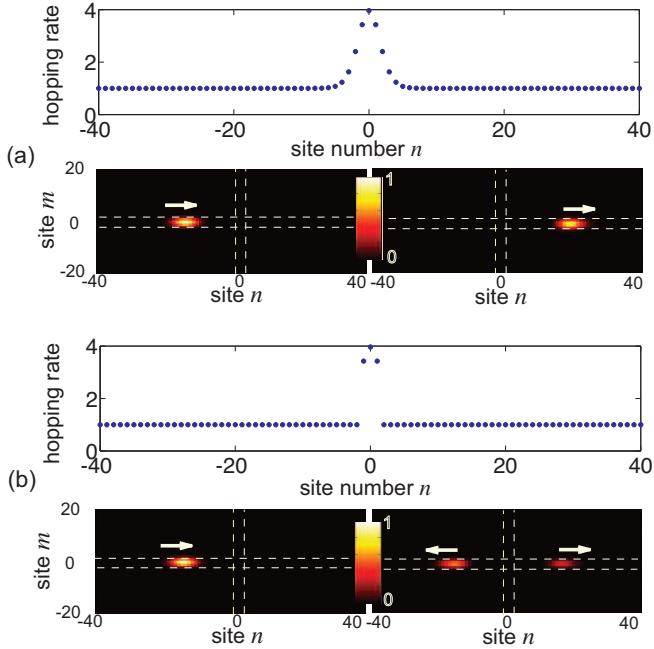}} \caption{ \small
(Color online)  Scattering of a Gaussian wave packet at the crossing between two guides in a square lattice of resonators. (a) Upper panel: behavior of the hopping rate $V_n=W_n$, normalized to the asymptotic value $\kappa$, as given by Eq.(6), corresponding to transparent crossing. Lower panels: two snapshots of $|c_{n,m}|^2$ (in arbitrary units) before (left panel) and after (right panel) the crossing. Arrows indicate the direction of propagation, whereas the straight dashed lines schematically depict the defective (guiding) regions. (b) Same as (a), but for a modified hopping rate $V_n=W_n$, leading to non-transparent intersection. Results are obtained by numerical simulations of Eq.(5) with the initial condition $c_{n,m}(0) \propto \exp[-i \pi n/2-(n+15)^2/25-m^2/4]$.}
\end{figure}

SUSY can also offer the possibility to design transparent intersections for discretized light \cite{S5,S8}. Let us consider, as an example, a square lattice of coupled resonators \cite{Fan1} with the same resonance frequency $\omega_R$ and with non-uniform hopping rates, as schematically shown in Fig.1(b). Indicating by $c_{n,m}$ the field amplitude in the resonator at lattice site $(n,m)$, with horizontal and vertical indices $n$ and $m$,  in the geometric setting of Fig.1(b) the following coupled-mode equations hold \cite{Fan1}
\begin{eqnarray}
i \frac{d c_{n,m}}{dt} & = & \omega_R  c_{n,m}+V_{n} c_{n-1,m}+V_{n+1}c_{n+1,m} \nonumber \\
& + & W_{m} c_{n,m-1}+W_{m+1}c_{n,m+1}
\end{eqnarray}
where $V_n$ is the hopping rate between resonators at site $(n,m)$ and $(n-1,m)$, whereas  $W_m$ is the hopping rate between resonators at site $(n,m-1)$ and $(n,m)$. We typically assume that inhomogeneities in the hopping rates are localized near $n=0$ and $m=0$, i.e. $V_n, W_n \rightarrow \kappa$ as $n \rightarrow \infty$, where $\kappa$ si the hopping rate of the homogeneous lattice. 
Like for the continuous Helmholtz equation (1), for the chosen functional dependence of hopping rates Eq.(5) is separable, i.e. $c_{n,m}(t)=F_n(t)G_m(t)$, leading to two 1D discrete Schr\"{o}dinger equations for $F_n$ and $G_n$. In particular, the defects of $V_n$ and $W_m$ near $n=0$ and $m=0$ can sustain bound propagative modes along the $n$ and $m$ directions, similar to the guided channels S1 and S2 in Fig.1(a). Interestingly, application of SUSY to the discrete Schr\"{o}dinger equations for $F_n$ and $G_m$ can be exploited to design reflectionless  defects \cite{S5}. For example, by assuming $V_n=\kappa Y_n(N,\sigma_1,\alpha_1)$ and 
$W_m=\kappa Y_m (M,\sigma_2,\alpha_2)$, where $Y_n(N,\sigma,\alpha)= $
\begin{equation}
\sqrt{ \frac{ {\rm{cosh}} [\sigma (n-\alpha) ] {\rm{cosh}} [ \sigma (n-\alpha -2N-1) ]} { {\rm{cosh}}[\sigma(n-\alpha -N)] {\rm{cosh}}[\sigma(n-\alpha -N-1)] }  }
\end{equation}
and $ \sigma$, $\alpha$ are arbitrary real parameters, one obtains transparent crossing of two guides along the $n$ and $m$ axes, sustaining $2N$ and $2M$ propagative modes along the two directions. Since the hopping rate is determined by evanescent tunneling of photons between resonators, hopping rate tailoring can be readily  obtained by a judicious control of the resonator distances. Transparency of  a special class of defects of the kind described by Eq.(6) has been recently proposed and demonstrated in 1D lattices in Refs.\cite{Szam1,Szam2}.  An example of transparent intersection is shown in Fig.5. Figure 5(a) shows the reflectionless propagation of a Gaussian wave packet along one of the two SUSY-synthesized defect waveguides for $V_n=W_n=\kappa Y_n(2,0.6,3.5)$, corresponding to multimode waveguides sustaining 4 modes. Broadening of the transmitted wave packet is visible, which is due to both multimode excitation and mode dispersion of the guide. For comparison, in Fig.5(b) the scattering of the same Gaussian wave packet is depicted for a different choice of the defects, clearly showing strong back reflection. \par
To conclude, broadband transparent intersections
between guiding structures in optical networks can be synthesized by application of SUSY. The present analysis is expected to be of interest in the design of high-density on-chip optical components, and can stimulate further studies. For example, with the application of SUSY to guiding structures with gain and loss, described by non-Hermitian Hamiltonians, one could design transparent intersection of active waveguides, i.e.  optical amplifiers. Moreover, extensions of SUSY to the 2D Helmholtz equation in the non-separable case, based on Moutard transform \cite{Mou}, could provide further design tools.  

\par

\newpage

\footnotesize {\bf References with full titles}\\
\\
1. J.B. Pendry, D. Schurig, and D.R. Smith, "Controlling electromagnetic fields", Science
{\bf 312}, 1780 (2006).\\
2. U. Leonhardt, "Optical conformal mapping", Science  {\bf 312}, 1777 (2006).\\
3.  D. Schurig, J.J. Mock, B.J. Justice, S.A. Cummer, J.B. Pendry, A. F. Starr,
and D. R. Smith, "Metamaterial electromagnetic cloak at microwave frequencies", Science {\bf 314},  977 (2006).\\
4. M. Rahm, D. Schurig, D.A. Roberts, S.A. Cummer, D.R. Smith, and J.B. Pendry,
"Design of electromagnetic cloaks and concentrators using form-invariant coordinate
transformations of Maxwell's equations",  	Photon. Nanostruct.: Fundam. Appl. {\bf 6}, 87 (2008).\\
5 E.E. Narimanov and A.V. Kildishev, "Optical black hole: broadband omnidirectional light
absorber",  Appl. Phys. Lett. {\bf 95}, 041106 (2009).\\
6. Y. Lai, J. Ng, H.Y. Chen, D.Z. Han, J.J. Xiao, Z.-Q. Zhang,
and C.T. Chan,"Illusion Optics: The Optical Transformation of an Object into Another Object", 
Phys. Rev. Lett. {\bf 102}, 253902 (2009).\\
7. Q. Cheng, K. Wu, Y. Shi, H. Wang, and G.P. Wang, "Directionally Hiding Objects and
Creating Illusions at Visible Wavelengths by Holography", Sci. Rep. {\bf 3}, 1974 (2013).\\
8. M.-A. Miri, M. Heinrich, R. El-Ganainy, and D. N. Christodoulides, "Supersymmetric Optical Structures," Phys. Rev. Lett. {\bf 110},
233902 (2013).\\
9. M.-A. Miri, M. Heinrich, and D. N. Christodoulides, "Supersymmetric transformation optics," Optica {\bf 1}, 89 (2014).\\
10. H. P. Laba and V. M. Tkachuk, "Quantum-mechanical analogy and supersymmetry of electromagnetic wave modes in planar
waveguides," Phys. Rev. A {\bf 89}, 033826 (2014).\\
11.  M. Heinrich, M.-A. Miri, S. St\"{u}tzer, R. El-Ganainly, S. Nolte, A. Szameit, and D. N. Christodoulides, "Supersymmetric mode
converters," Nature Commun. {\bf 5}, 3698 (2014).\\
12. S. Longhi, "Invisibility in non-Hermitian tight-binding lattices," Phys. Rev. A {\bf 82}, 032111 (2010).\\
13. S. Longhi and G. Della Valle, "Transparency at the interface between two isospectral crystals," EPL {\bf 102}, 40008 (2013).\\
14. S. Longhi and G. Della Valle, "Invisible defects in complex crystals," Ann. Phys. {\bf 334}, 3546 (2013).\\
15.M. Heinrich, M.-A. Miri, S. St\"{u}tzer, S. Nolte, D.N. Christodoulides, and A. Szameit, "Observation of supersymmetric scattering
in photonic lattices", Opt. Lett. {\bf 39}, 6130 (2014).\\
16. H. Liu, H. Tam, P.K.A. Wai, and E. Pun, "Low-loss waveguide crossing using a multimode interference structure,"
Opt. Commun. {\bf 241}, 99 (2004).\\
17. S.-H. Kim, G. Cong, H. Kawashima, T. Hasama,
and H. Ishikawa, "Tilted MMI crossings based on silicon
wire waveguide", Opt. Express {\bf 22}, 2545 (2014).\\
18. S.G. Johnson, C. Manolatou, S. Fan, P.R. Villeneuve, J.D. Joannopoulos, and H.A. Haus, "Elimination of cross talk in waveguide intersections", Opt. Lett. {\bf 23}, 1855 (1998).\\
19. T. Fukazawa, T. Hirano, F. Ohno, and T. Baba, "Low loss intersection of Si photonic wire waveguides," Jpn. J.
Appl. Phys. {\bf 43}, 646 (2004).\\
20. F. Shinobu, Y. Arita, and T. Baba, "Low-loss simple waveguide intersection in silicon photonics," Electron. Lett.
{\bf 46}, 1149 (2010).\\
21. W. Bogaerts, P. Dumon, D. V. Thourhout, and R. Baets, "Low-loss, low-cross-talk waveguide crossings for
silicon-on-insulator on nanophotonic waveguides," Opt. Lett. {\bf 32}, 2801 (2007).\\
22. Q. Wu, J.P. Turpin, and D.H. Werner, "Integrated photonic systems based on transformation optics enabled gradient index devices", Light: Science \& Applications {\bf 1}, e38 (2012).\\
23. J. Li, D.A. Fattal, and R.G. Beausoleil, "Crosstalk-free design for the intersection of two dielectric waveguides", Opt. Express {\bf 17}, 7717 (2009).\\
24. E. Sh. Gutshabash, "Moutard transformation and its application to some physical problems. I. The case of two independent variables", 
J. Math. Sciences (New York) {\bf 192}, 57 (2013).\\
25. L. D. Landau and E. M. Lifshitz, {\it Quantum Mechanics} (Pergamon, Oxford,
1965), 2nd ed., Secs. 23 and 25.\\
26. J. Valentine, J. Li, T. Zentgraf, G. Bartal, and X. Zhang, "An optical cloak made of dielectrics",
Nat. Mater. {\bf 8}, 568 (2009).\\
27. J. Hunt, R. Tyler, S. Dhar, Y.J. Tsai, P. Bowen, S. Larouche, N.M. Jokerst, and D.R. Smith, "Planar, flattened Luneburg lens at
infrared wavelengths",  Opt. Express {\bf 20}, 1706 (2012).\\
28. S.K. Yao, D.B. Anderson, R.R. August, B.R. Youmans, and C.M. Oania, "Guided-wave optical thin-film
Luneburg lenses: fabrication technique and properties", Appl. Opt. {\bf 18}, 4067(1979).\\
29. K. Fang, Z. Yu, and S. Fan, "Realizing Effective Magnetic Field for Photons by Controlling the Phase of Dynamic
Modulation", Nat. Photonics {\bf 6}, 782 (2012).\\
30. A.A. Sukhorukov, "Reflectionless potentials and cavities in waveguide arrays and coupled-resonator structures," Opt. Lett. {\bf 35},
989991 (2010).\\
31. A. Szameit, F. Dreisow, M. Heinrich, S. Nolte, and A. A. Sukhorukov, "Realization of Reflectionless Potentials in Photonic Lattices,"
Phys. Rev. Lett., {\bf 106}, 193903 (2011).

\end{document}